\documentclass[aip,jcp,amsmath,12pt]{revtex4-1}
\usepackage{graphicx}% Include figure files
\usepackage{dcolumn}% Align table columns on decimal point
\usepackage{bm}% bold math

\newcommand*{\beq}[0]{\begin{equation}}
\newcommand*{\eeq}[0]{\end{equation}}
\newcommand{\beqs}{\begin{eqnarray}}
\newcommand{\eeqs}{\end{eqnarray}}

\newcommand*{\eqn}[1]{Eqn.~\ref{eqn:#1}}
\newcommand*{\fig}[1]{Figure~\ref{fig:#1}}
\newcommand*{\Sec}[1]{Section~\ref{sec:#1}}

\newcommand*{\x}[0]{\times}
\newcommand*{\rv}[0]{\vec r}
\newcommand*{\xv}[0]{\vec x}
\newcommand*{\jv}[0]{\vec j}
\newcommand*{\VX}[0]{V_X}
\newcommand*{\VXp}[0]{V_{X'}}
\newcommand*{\aXX}[0]{a_{X\to X'}}
\newcommand*{\fXX}[0]{\partial V_{X\to X'}}
\newcommand*{\nXX}[0]{\hat{n}_{X\to X'}}
\newcommand*{\lXX}[0]{\ell_{X\to X'}}
\newcommand*{\tauXX}[0]{\tau_{X\to X'}}
\newcommand*{\JXX}[0]{J_{X\to X'}}
\newcommand*{\wXA}[0]{w^A(X)}
\newcommand*{\wXpA}[0]{w^A(X')}
\newcommand*{\grid}[0]{h}

\begin{document}

\title{Accurate and efficient algorithm for Bader charge integration}

\author{Min Yu}
\affiliation{Department of Physics, University of Illinois at
Urbana-Champaign, Urbana, IL 61801}
\author{Dallas R. Trinkle}
\email{dtrinkle@illinois.edu}
\affiliation{Department of Materials Science and Engineering, University of
Illinois at Urbana-Champaign, Urbana, IL 61801}
\date{\today}

\begin{abstract}
We propose an efficient, accurate method to integrate the basins of
attraction of a smooth function defined on a general discrete grid, and
apply it to the Bader charge partitioning for the electron charge density.
Starting with the evolution of trajectories in space following the gradient
of charge density, we derive an expression for the fraction of space
neighboring each grid point that flows to its neighbors.  This serves as
the basis to compute the fraction of each grid volume that belongs to a
basin (Bader volume), and as a weight for the discrete integration of
functions over the Bader volume.  Compared with other grid-based
algorithms, our approach is robust, more computationally efficient with
linear computational effort, accurate, and has quadratic convergence.
Moreover, it is straightforward to extend to non-uniform grids, such as
from a mesh-refinement approach, and can be used to both identify basins of
attraction of fixed points and integrate functions over the basins.
\end{abstract}
\pacs{}
\maketitle

\section{Introduction}
Based on density functional theory (DFT)\cite{Theor:KS} calculations,
decomposing the charge or the energy of a material into contributions from
individual atoms can provide new information for material properties.
Bader's ``atoms in molecules'' theory provides an example of a partitioning
based on the charge density, and following the gradient at a particular
point in space to the location of a charge density maximum centered at an
atom---defining basins of attraction of fixed points of the charge density.
Bader defines the atomic charges and well-defined kinetic energies as
integrals over these Bader volumes,\cite{Bader} $\Omega_{\rho}$.  Each
Bader volume contains a single electron density maximum, and is separated
from other volumes by a zero flux surface of the gradients of the electron
density, $\nabla \rho(\rv) \cdot \hat{n}=0$. Here, $\rho(\rv)$ is the
electron density, and $\hat{n}$ is the unit vector perpendicular to the
dividing surface at any surface point $\rv\in\partial\Omega_{\rho}$.  Each
volume $\Omega_{\rho}$ is defined by a set of points where following a
trajectory of maximizing $\rho$ reaches the same unique maximum (fixed
point).  In practical numerical calculations, where the charge density is
defined on a discrete grid of points in real space, it is very challenging
to have an accurate determination of a zero flux surface.

Different approaches for condensed, periodic systems have relied
on analytic expressions of the density\cite{Bader1981,Morphy1996}
or discretizing the charge density
trajectories\cite{Bader:OCTREE2003,Uberuaga1999,Bader:onGridMethod,
Bader:nearGridMethod,Otero2010}.  Early algorithms were based on
the electron density calculated from analytical wavefunctions of
small molecules, and integration along the gradient paths.  Most
current developments are based on a grid of electron density, which
is important for DFT calculation and also applicable to analytical
density function of small molecules.  One octal tree
algorithm\cite{Bader:OCTREE2003} uses a recursive cube subdivision
to find the atomic basins robustly, but practically is not applicable
to complicated topologies due to huge computational cost.  The
``elastic sheet" method\cite{Uberuaga1999} defines a series of
fictitious particles which gives a discrete representation of
zero-flux surface. Particles are relaxed according to the gradients
of charge density and interparticle forces. This method will not
work for complex surface with sharp cusps or points.  Recently,
Henkelman \textit{et al.} developed an on-grid
method\cite{Bader:onGridMethod} to divide an electron density grid
into Bader volumes.  This method can be applied to the DFT calculations
of large molecules or materials.  They discretize the trajectory
to lie on the grid, ending at the local maximal point of the electron
density.  The points along each trajectory are assigned to the atom
closest to the end point.  Although this method is robust, and
scales linearly with the grid size, it introduces a lattice bias
caused by the fact that ascent trajectories are constrained to the
grid points.  The near-grid method\cite{Bader:nearGridMethod}
improves this by accumulating a correction vector---the difference
between the discretized trajectory and the true trajectory---at
each step.  When the correction vector is sufficiently large, the
discrete trajectory is corrected to a neighboring grid point.  This
method corrects the lattice bias, and also scales linearly with
respect to the size of grids.  However, both grid trajectory methods
require iteration to self-consistency in volume assignments.  Also,
the integration error scales linearly with the grid spacing, so
very fine grids are required in numerical calculations to provide
the correct Bader volume, reducing its applicability for accurate
calculations in a large system.  Lastly, a new algorithm uses a
``divide and conquer'' adaptive approach with tetrahedra; tetrahedra
are continuously divided at the boundaries of Bader volumes, with
the weight of each tetrahedra given by the number of vertices that
belong to each volume\cite{Otero2010}.  Such an approach retains
linear scaling with the grid spacing, but requires mesh refinement
near boundaries to deal with the linear convergence of the error
with the grid spacing.

\begin{figure}[!ht]
\begin{center}
  \includegraphics[width=3in]{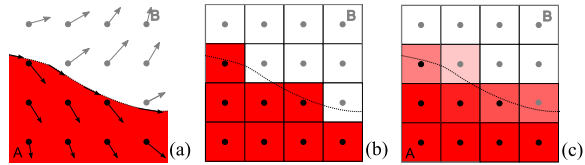}
\end{center}
\caption[A schematic illustration of the zero flux surface, the
near-grid algorithm, and weighted integration]{A schematic illustration of
the zero flux surface, the near-grid algorithm, and weighted
integration. The zero flux dividing surface separate volumes A and B, where
arrows denote charge density gradients (a). The normal component is zero
for any point on the surface $\nabla \rho \cdot \hat{n}=0$. The near-grid
method\cite{Bader:nearGridMethod} gives grid-based partition (b), however
energy density integration based on this grid-based partition would cause
integration error due to finite grid sizes. A weight function (c)
representing volume fractions of the cell of each grid point is introduced
to reduce the error due to a finite grid.}
\label{fig:SchematicModel}
\end{figure}

\fig{SchematicModel} illustrates the partition of the real
space into two Bader volumes A and B by a zero flux dividing surface.  The
component of $\nabla\rho$ along the surface normal $\hat{n}$ is zero for
any point on the surface $\partial A$ or $\partial B$.  A grid-based
partition algorithm, such as the near-grid method, divides space into
volume surrounding around each grid point, and assigns each grid volume to
a particular Bader volume.  Even though grid points may be assigned to
Bader volumes correctly, the density integration based on the grid-based
partition would bring in numerical integration error that scales linearly
with the grid spacing.  Introducing a ``weight'' integrand representing the
fraction of grid volume that belongs to a particular Bader volume smooths
out the grid-based partition, and improves the integration accuracy and
scales quadratically with the grid spacing.  The atomic contribution is
neither 1 nor 0 at the dividing surfaces, but fractional.  In
\fig{SchematicModel}, red represents a weight of 1 to atom A for grid points
closer to atom A, and transitions to white for a weight of 0 for grid
points away from atom A.

The grid-based weight representing volume fractions of each grid volume
assigned to different atoms, and gives a more accurate integrand for the
integration of either charge density or of kinetic energy over the Bader
volumes.  The weight is computed from the total integrated flux of
trajectories in a grid volume to neighboring grid volumes.  The algorithm
is robust, efficient with linear computing time in the number of grid
points, and more accurate than other grid-based algorithm.  Surprisingly,
it combines both better error scaling---quadratic in the grid spacing---and
improved computational efficiency.  Moreover, it is straightforward to
apply to nonuniform grids, such as would result from an adaptive
mesh-refinement approach.  In \Sec{method}, we derive the algorithm to
constructing a grid-based weight to perform numerical integrals over Bader
volumes.  \Sec{tests} presents examples including three dimensional charge
density from three Gaussian functions in FCC cell, TiO$_2$ bulk, and NaCl
crystal.  Finally, we show the improved computational efficiency in
\Sec{comput}.  The end result is a simple, extendable, computationally
efficient algorithm with quadratic integration error.

\section{The Weight Method}
\label{sec:method}

The Bader partitioning of space defines volumes by the endpoint of a
trajectory following the gradient flow of the charge density, $\nabla\rho$.
We assume that $\rho$ has continuous first and second derivatives
throughout all space of interest, and has a set of discrete local maxima
(fixed points) $\xv_1$, $\xv_2$, etc., where $\nabla\rho=0$ and the matrix
$\nabla\nabla\rho$ is negative-definite.  The basin of attraction $A_n$, of
a fixed point $\xv_n$ is the set of points which flow to the fixed point
$\xv_n$ along the charge density gradient.  That is, for any point $\rv$,
we can integrate the trajectory given by $\dot\xv(t) =
\nabla\rho(\xv)$, with the initial condition $\xv(0)=\rv$, to find
$\lim_{t\to\infty}\xv(t)$.  Each trajectory will end at fixed point
$\xv_n$, and except for a set of points with zero volume in space, the
extremum is a local maximum; the basin of attraction $A_n$ are all points
$\rv$ whose trajectory $\lim_{t\to\infty}\xv(t)$ ends at $\xv_n$.  Note
also that if point $\rv_0\in A$, and the trajectory starting from $\rv_1$
reaches $\rv_0$ in a finite time $t$, then $\rv_1\in A$.  This set defines
a partitioning of space, where $A_n\cap A_m=\emptyset$ when $n\ne m$ and
$\cup_n A_n=\Omega$.  Finally, each basin $A_n$ is such that wherever the
normal $\hat n$ to the bounding surface $\partial A_n$ is well-defined,
$\hat n\cdot\nabla\rho = 0$.  If $\rho$ is the charge density, then $A_n$
are the Bader volumes; but this definition is applicable to any
sufficiently smooth function with a discrete set of local maxima.  As the
definition of the basins $A_n$ derives from trajectories, it is not
possible in general to determine if two neighboring points $\rv$ and $\rv'$
belong to the same or different basins based only on local information.

\begin{figure}[!ht]
 \begin{center}
  \includegraphics[width=2in]{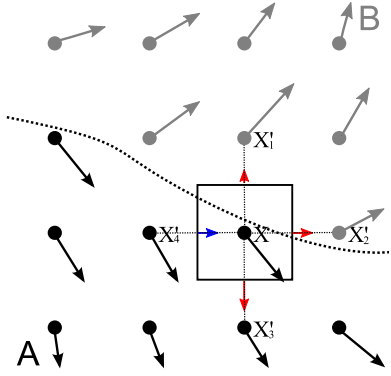}
   \caption[Schematic illustration of the weight method.]{Schematic
illustration of the weight method. The volume of the cell of a grid point
flows to its neighbors with larger charge density magnitude. Flowing flux
is shown as directional map, either flowing from $X$ to $X'$ as red arrows
or flowing from $X'$ to $X$ as blue arrows.}
 \label{fig:WeightMethod}
 \end{center}
\end{figure}

\fig{WeightMethod} shows the reformulation for an approximate fractional
partitioning of real-valued function evaluated at a set of discrete points,
$X$.  The grid points $X$ partition space into Voronoi
polyhedra\cite{Voronoi1907} $\VX$ covering each grid point $X$, where a
point in space $\rv$ belongs to the volume $\VX$ if $X$ is the closest
point in Cartesian space to $\rv$.  Each polyhedra is defined by the
nearest neighboring points $X'$ that are a distance $\lXX$ away; the
Voronoi polyhedron at $X$ has facets $\fXX$ with normal $\nXX$ pointing
from $X$ to $X'$ and area $\aXX$.  Moreover, the facet is at the midpoint
between $X$ and $X'$.  Our goal is to define for each grid point $X$, a
``weight'' $\wXA$ between 0 and 1 such that $\sum_A \wXA = 1$ for all $X$,
and the discrete approximation to the integral over the basin $A$
\beq
\int_{A} d^3r f(\rv) \approx \sum_X \VX \wXA f(X)
\label{eqn:integrate}
\eeq
converges quadratically in the grid spacing for smooth functions $f(\rv)$.
The weight, in this case, is the fraction of points in $\VX$ whose
trajectory ends in the basin $A$.  Note that if the points $X$ form a
regular periodic grid, the Voronoi volumes, facet areas, and neighbor
distances need only be computed for the Wigner-Seitz cell around a grid
point.

To transition from the continuum definition of spatial partitioning to our
Voronoi partitioned definition, we introduce the continuum probability
density for our trajectories, $P(\rv, t)$.  From the trajectory equation,
the probability flux at any point and time is
$\jv(\rv,t)=P(\rv,t)\nabla\rho(\rv)$.  Then, the probability distribution
evolves in time according to a continuity equation
\beq
\frac{\partial P(\rv,t)}{\partial t} + \nabla\cdot(P(\rv,t)\nabla\rho(\rv))
= 0.
\eeq
This equation represents the combined evolution of a distribution of points
in space; we use it to determine how the points in $\VX$ distribute to
neighboring volumes $\VXp$.  Define the volume probability
\beq
P_X(t) = \VX^{-1} \int_{\VX} d^3r P(\rv, t);
\eeq
then the evolution from the initial condition
\beq
P(\rv, 0) = \begin{cases}
1&: \rv\in \VX\\
0&: \rv\notin \VX
\end{cases}
\eeq
is given by
\beq
\begin{split}
\frac{dP_X(t)}{dt} &= - \VX^{-1} \int_{\VX} d^3r
\nabla\cdot(P(\rv,t)\nabla\rho(\rv))\\
&= -\VX^{-1}\sum_{X'} \int_{\fXX} P(\rv, t) \nabla\rho\cdot\nXX d^2r\\
&\approx -\VX^{-1} P_X(t) \sum_{X'} \int_{\fXX} \nabla\rho\cdot\nXX d^2r\\
&\approx -P_X(t) \sum_{X'} \frac{\aXX}{\VX}\cdot\frac{R(\rho_{X'}-\rho_X)}{\lXX}\\
&\equiv -P_X(t) \sum_{X'} \tauXX
\end{split}
\label{eqn:ProbEvol}
\eeq
where $R(u)=u\theta(u)$ is the ramp function, so that $\tauXX\ge0$ and is
zero when $\rho_{X'} < \rho_X$; this is a consequence of our initial
conditions where $P(r,0)$ is only nonzero in the interior of $\VX$.  The
first approximation ignores spatial variation of $P(r,t)$ through the
volume $\VX$ (an error linear in the grid spacing), and the second
approximation ignores spatial variation of $\nabla\rho$ along a facet
$\fXX$, and approximates the gradient at the midpoint between $X$ and $X'$
with the finite difference value (also with an error that is linear in the
grid spacing).  The solution to \eqn{ProbEvol} is $P_X(t) =
\exp(-t\sum_{X'} \tauXX)$.  For that solution, the time-integrated flux of
probability from $\VX$ to $\VXp$ through the facet $\fXX$ is
\beq
\begin{split}
\JXX &= \int_0^\infty dt \int_{\fXX} P(\rv,t)\nabla\rho\cdot\nXX d^2r\\
&\approx \int_0^\infty dt P_X(t) \int_{\fXX} \nabla\rho\cdot\nXX d^2r\\
&\approx \int_0^\infty dt P_X(t) \tauXX\\
% &= \frac{\tauXX}{\sum_{X'}\tauXX}\\
&= \frac{\aXX\lXX^{-1} R(\rho_{X'}-\rho_X)}{\sum_{X'}\aXX\lXX^{-1} R(\rho_{X'}-\rho_X)}
\end{split}
\label{eqn:flux}
\eeq
where we have used the same approximations as above.  This flux defines the
total fraction of points inside $\VX$ that transition to volume $\VXp$
through $\fXX$.  Note that $\sum_{X'} \JXX = 1$, unless $X$ is a local
(discrete) maxima, where $\rho_X > \rho_{X'}$ for all neighbors $X'$.
Finally, as the weight $\wXA$ represents the volume fraction of points in
volume $\VX$ whose trajectory ends inside basin $A$, then
\beq
\wXA = \sum_{X'} \JXX \wXpA.
\label{eqn:weight}
\eeq
Note that if for all $X'$ where $\rho(X')>\rho(X)$, $\sum_A \wXpA = 1$,
then as $\sum_{X'} \JXX = 1$, \eqn{weight} guarantees that $\sum_A \wXA =
1$.  Appendix~\ref{sec:quad} shows that the error in the weight of linear
order in the grid spacing produces a quadratic order error in the
integration.

Forward substitution solves \eqn{weight} after the grid points are sorted
from highest to lowest density $\rho(X)$.  Sequentially, each point $X$ is
either
\begin{enumerate}
\item A local maxima: $\rho(X)>\rho(X')$ for all neighbors $X'$.  This grid
point corresponds to a new basin $A$, and we assign $\wXA=1$.
\item An interior point: for all $X'$ where $\rho(X')>\rho(X)$, the weights
have been assigned and $\wXpA=1$ for the \textit{same} basin $A$.  Then
\eqn{weight} assigns $X$ to basin $A$ as well: $\wXA=1$.
\item A boundary point; with weights between 0 and 1 for multiple basins
assigned by \eqn{weight}.
\end{enumerate}
Then $\wXA$ is known from $\wXpA$ where $\rho(X')>\rho(X)$ for each basin
$A$ (as $\JXX\ne0$ only if $\rho(X')>\rho(X)$).  Note also that the weight
for a particular basin $A_n$ is assigned \textit{without} reference to any
other basin $A_m$; once the set of time-integrated fluxes $\JXX$ are known
and the densities sorted in descending order, the solution for each basin
is straightforward, and
\eqn{weight} is only needed on the boundary points.

This algorithm solves several issues with the near-grid method.  It
requires no self-consistency, which improves the computational scaling.
Moreover, the introduction of smooth functions that define the volume
fraction of points in each basin produces less error and faster convergence
with additional grid points.  The algorithm is also readily applicable to
non-uniform grids, such as an adaptive meshing scheme---it only requires
computation of the Voronoi volumes and facets for the grid points.  In one
dimension, \eqn{integrate} has quadratic convergence in the grid spacing
(c.f. Appendix~\ref{sec:quad}); we now demonstrate the quadratic
convergence and improved integration accuracy for three dimensional
problems.

\section{Evaluation of numerical convergence}
\label{sec:tests}
One determination of the accuracy of Bader volume integration is the
vanishing of the volume integration of the Laplacian of charge density
$\nabla^{2}\rho(\mathbf{r})$.  The non-zero value of the Laplacian of
charge density integration within each Bader volume is our atomic
integration error, and can be used as an estimate of the error of the
integration of the kinetic energy.  We construct the zero flux surface of
the gradients of charge density and evaluate the integration error with
both the weight- and near-grid methods for several cases.  First, we
consider an analytic charge density with known boundaries in an orthogonal
and a non-orthogonal cell.  Next, we calculate real systems: an ionic
compound, and a semiconductor.  We also evaluate the Bader charge of Na
atom in NaCl crystal by integrating the charge density within Bader volume,
and compare the convergence with the near-grid method.

\subsection{Gaussian densities}

\begin{figure}[!ht]
\begin{center}
  \includegraphics[width=2in]{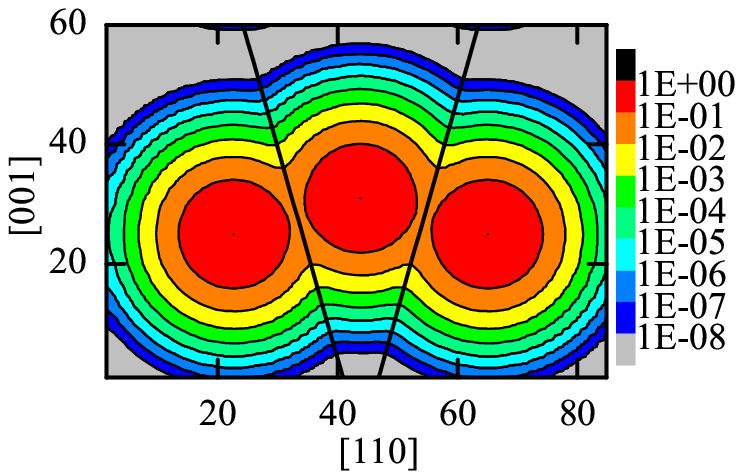}\\
  \includegraphics[width=2in]{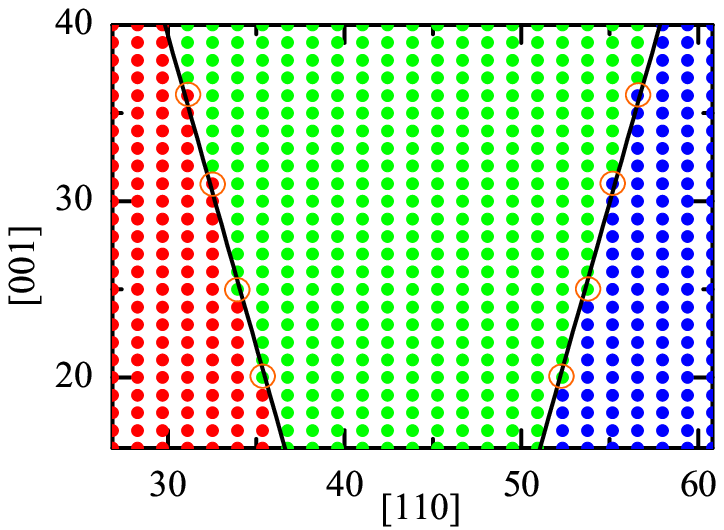}\\
  \includegraphics[width=2in]{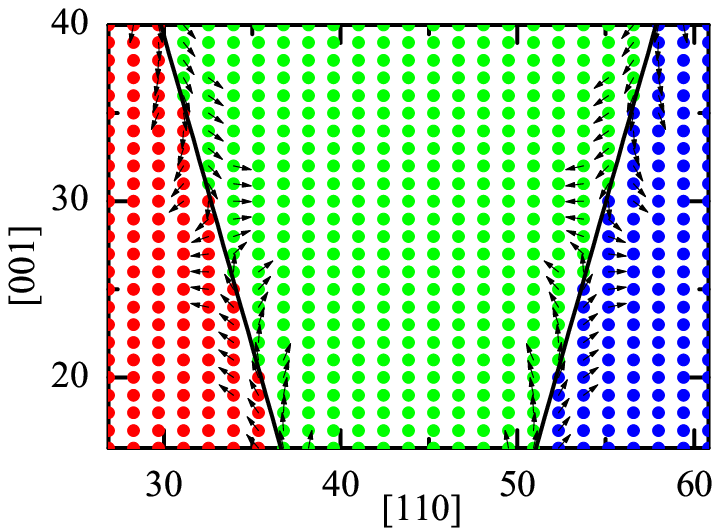}
\end{center}
\caption[Charge density distribution constructed from three Gaussian
functions]{Charge density distribution constructed from three Gaussian
functions, and basin identification with near-grid and the weight method.
The grid size is $N=60$, and the charge density is shown on $(1\bar{1}0)$
plane; this is coplanar with the centers of the three Gaussian functions.
The true dividing surfaces are indicated by two black lines due to
symmetry.  The basin assignment of grid points is given: red dots for ion
I, green dots for ion II, and blue dots for ion III; for the weight method,
a single color is assigned to the maximum weight at each point.  Basin
assignment from the near-grid method is given in the middle panel; basins
with maximal weight on every grid points from the weight method are
indicated in the right panel.  Orange circles in the middle panel indicate
the grid points misassigned by the near-grid method and corrected by the
weight method.  Arrows in the bottom panel denote the directions of the
gradients of charge density, which can be used to verify the correctness of
basin assignment.}
\label{fig:3DGaussianCharge}
\end{figure}

\fig{3DGaussianCharge} shows an example of misassignment of the grid points
to basins from the near-grid method.  Misassignment occurs for the grid
points close to the dividing surfaces with the gradients of charge density
almost parallel to the surfaces. In this example, a three dimensional model
charge density is constructed from three Gaussian functions in simple cubic
unit cell, $\rho(\rv)=\sum_{i=1,3}e^{(-\rv-\rv_i)^2/W^2}$. The $\rv_i$ are
$(0.25N,0.25N,0.4N)$, $(0.5N,0.5N,0.5N)$, and $(0.75N,0.75N,0.4N)$, with
width $W=N/10$. \fig{3DGaussianCharge} shows the charge density
distribution on $(1\bar{1}0)$ plane. Due to the symmetry of charge density
distribution, the true dividing surfaces along charge density saddle points
are known analytically and shown as two black lines on $(1\bar{1}0)$
plane. The grid points marked by orange circles are assigned to the wrong
basins by the near-grid method, different from the partition of spatial
points by true dividing surfaces.  The gradients of charge density shown in
arrows for these misassigned surface points have small normal components,
and we believe this is the cause of the misassignment.

\begin{figure}[!ht]
\begin{center}
  \includegraphics[width=3in]{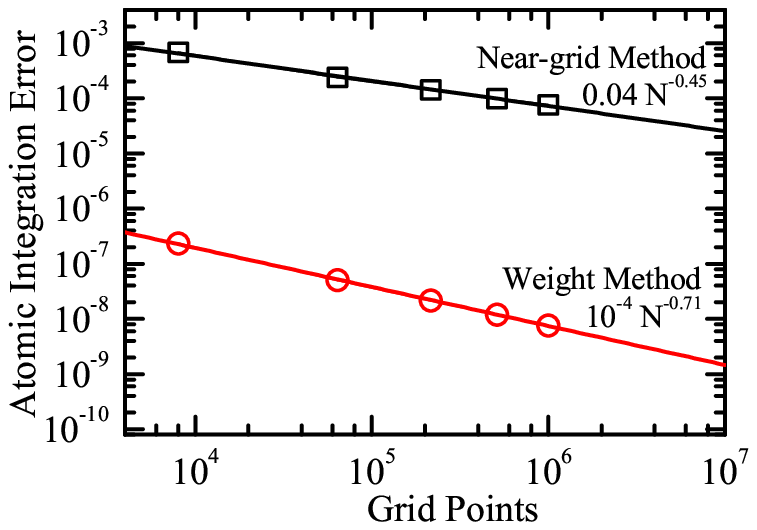}
\end{center}
\caption[Comparison of the near-grid method and the weight method on atomic
integration errors in a FCC cell.]{Comparison of the near-grid method and
the weight method on atomic integration errors in a FCC cell. The maximal
atomic volume integrations of the Laplacian of charge density within Bader
volumes using the near-grid method and the weight method are denoted by
squares and circles, respectively.  We calculate charge density grids
ranging from $20^3$ points to $100^3$.  Our algorithm gives atomic
integration errors three orders of magnitude lower than the near-grid
method, and converges faster than the near-grid method.}
\label{fig:3DGaussian-FCC}
\end{figure}

To test integration accuracy beyond simple cubic lattices, we map this
model charge density onto a FCC unit cell shown in
\fig{3DGaussian-FCC}. The three dimensional model charge density is
constructed from three Gaussian functions,
$\rho(\mathbf{r})=\sum_{i=1,3}e^{(-\mathbf{r}-\mathbf{r}_i)^2/W^2}$.  The
$\mathbf{r}_i$ are located at $(0.25N,0.25N,0.4N)$; $(0.5N,0.5N,0.5N)$;
$(0.75N,0.75N,0.4N)$ where $N^3$ is the number of grid points in the FCC
unit cell and $W=N/10$.  We vary $N$ from 20 to 100.  The Voronoi cell of
FCC lattice has 12 neighbors, where all facets have the same area. The
atomic weights on every grid represents the fraction of Voronoi volume of
that grid point flowing to specific atom through its neighbors.  By
calculating on a set of grid sizes, one obtains the maximal atomic
integration errors from the near-grid method and the weight method.

\fig{3DGaussian-FCC} shows a reduction in error of three orders of
magnitude from the near-grid method. Fitting data to a non-linear function
$y=a\,N^{-r}$ gives a convergence rate of 0.71 for the weight method, and
0.45 for the near-grid method.  The exponent of 0.71 is close to the 2/3
expected for quadratic convergence, and 0.45 is close to the 1/3 expected
for linear convergence.  The weight method has both better absolute error
and converges faster than the near-grid method; in addition, there is no
crossover point at large grid spacing where near-grid has smaller errors.

% \subsection{TiO$_2$ bulk}
\subsection{Titania bulk}
\begin{figure}[!ht]
\begin{center}
  \includegraphics[width=3in]{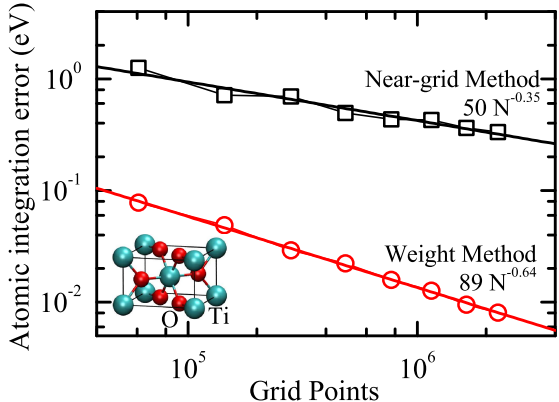}
\end{center}
\caption[Maximal atomic integration error on rutile TiO$_2$ with respect to
the charge density grids.] {Maximal atomic integration error on rutile
TiO$_2$ with respect to the charge density grids. A set of charger density
grids ranging from $45\x45\x30$ points to $150\x150\x100$ points are
calculated. The weight method reports maximal atomic integration error at
least one order of magnitude smaller than the near-grid method. The weight
method is practically useful for calculation of small grid size.}
\label{fig:TiO2_Tc}
\end{figure}

For a real charge density, we perform DFT calculations on TiO$_2$ bulk by
use of the projector augmented wave (PAW)\cite{Blochl1994} method, the GGA
with PBE functional\cite{Theor:PBE} for the exchange-correlation energy.
Density-functional theory calculations are performed with
\textsc{vasp}\cite{Kresse93,Kresse96b} using a plane-wave basis
with the projector augmented-wave (PAW) method,\cite{Blochl1994} with
potentials generated by Kresse.\cite{Kresse1999} Atomic configurations for
Ti and O are [Ne]$3s^{2}3p^{6}4s^{2}3d^{2}$ with cutoff radius 1.22\AA, and
[He]$2s^{2}2p^{4}$ with cutoff radius 0.58\AA, respectively.  We use a
plane-wave basis set with cut-off energy of 900eV.  The tetragonal unit
cell of rutile TiO$_2$ (see \fig{TiO2_Tc}) contains two Ti atoms and four O
atoms. Monkhorst-Pack k-point method with $4\x 4\x 6$ k-points for
six-atom cell is used for Brillouin-zone integration with a Gaussian
smearing of 0.1eV for electronic occupancies.  Theoretically optimized
lattice constant are $a=4.649\text{\AA}$, $c=2.970\text{\AA}$, $u=0.305$
agreeing with experimental lattice constants of $a=4.584\text{\AA}$,
$c=2.953\text{\AA}$, $u=0.305$\cite{TiO2:SSReport}. A set of charge
density grids ranging from $45\x45\x30$, $60\x60\x40$,
$75\x75\x50$, $90\x90\x60$, $120\x120\x80$
points to $150\x150\x100$ are calculated.  For the energy cutoff
of 900eV, a grid of $45\x45\x30$ is required to eliminate
wrap-around errors, and is the minimum size used by an accurate
\textsc{vasp} calculation.

\fig{TiO2_Tc} shows maximal atomic integration errors as a function of grid
sizes.  The weight method gives maximal atomic integration error one order
of magnitude lower than the near-grid method systematically.  The atomic
integration error larger than 1.0eV on the minimal grid size $45\x45\x30$
from the near-grid method is unacceptably large.  Again, the convergence
rate of the error goes as $\sim2/3$ for the weight method---corresponding
to quadratic convergence---and $\sim1/3$ for the near-grid
method---corresponding to linear convergence.  Both the improved error and
faster convergence allows for more accurate density integration with fewer
grid points than near-grid.

\subsection{NaCl crystal}
In this example, we evaluate the Bader charge (valence electron density
integration within Bader volume) of Na atom in NaCl crystal by integrating
the charge density within Bader volume, and compare the value with the
near-grid method.  We perform DFT calculations by use of the PAW method,
the GGA with PW91 functional\cite{Theor:PW91} for the exchange-correlation
energy. Atomic configurations for Na and Cl are [He]$2s^{2}2p^{6}3s^{1}$
with cutoff radius $0.77\,\text{\AA}$, and [Ne]$3s^{2}3p^{5}$ with cutoff
radius $1.00\,\text{\AA}$, respectively. A plane-wave basis set with
cut-off energy of 500eV is applied. The NaCl unit cell contains 4 Na
atoms and 4 Cl atoms.  Monkhorst-Pack k-point method with $3\x3\x3$
k-points for eight-atom cell is used for Brillouin-zone integration with a
Gaussian smearing of 0.2eV for electronic occupancies.  The optimized
lattice constant of 5.67\AA agrees with the experimental lattice constant
of 5.64\AA. A set of charge density grids of $60^3$, $80^3$, $100^3$,
$120^3$, to $180^3$ points are calculated.

\begin{figure}[!ht]
 \begin{center}
  \includegraphics[width=3in]{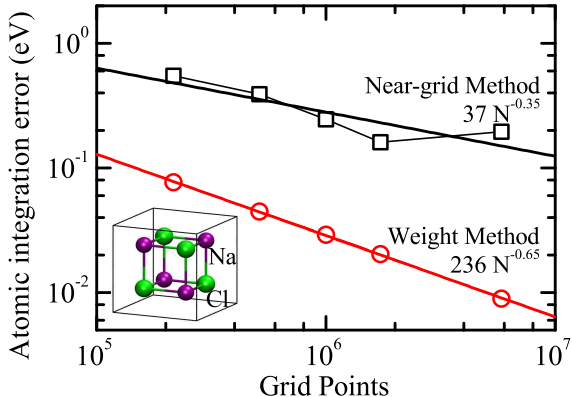}
   \caption[Comparison of the near-grid method and the weight method for maximal atomic integration error of NaCl crystal.]{Comparison of the near-grid method and the weight method for maximal atomic integration error of NaCl crystal. A set of charge density grids ranging from $60^3$ points to $180^3$ points are calculated. Comparing to the near-grid method, the weight method reduce the integration error remarkably.}
 \label{fig:NaCl_TcError}
 \end{center}
\end{figure}

\fig{NaCl_TcError} shows the maximal atomic integration error as a function
of various grid sizes.  The weight method again shows maximal atomic
integration error at least one order of magnitude lower than the near-grid
method systematically.  The scaling of the error goes as the $\sim 2/3$
power for the weight method, showing continued quadratic convergence, while
the near-grid method error scales as the $\sim 1/3$ power, which is linear
convergence.

\begin{figure}[!ht]
 \begin{center}
  \includegraphics[width=3in]{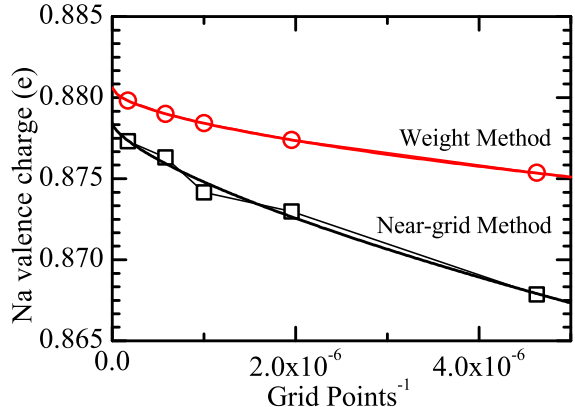}
   \caption[Comparison of the near-grid method and the weight method on convergence of Bader charges of Na in NaCl crystal.]{Comparison of the near-grid method and the weight method on convergence of Bader charges of Na in NaCl crystal. The Bader charge of Na is calculated for a set of density grids ranging from $60^3$ points to $180^3$ points. Both methods give monotonic, and smooth convergence.}
 \label{fig:NaCl_Charge}
 \end{center}
\end{figure}

\fig{NaCl_Charge} shows that Bader charge of Na atom evaluated on various
charge grids. The weight method computes a Bader charge of Na atom slightly
larger than the near-grid method.  Fitting the data to $\rho =
\rho_{0}+\frac{C}{N_{grid}^{\alpha}}$, we find converged Bader charge
values of $0.878$e, $0.881$e, for the near-grid method and the weight
method, respectively.  We believe this is due to a systematic misassignment
for the near-grid method, as shown for the Gaussian charge density case.
This suggests that the misassignment may not be improved by increasing the
density of grid points in the near-grid method.  This suggests that a
``divide and conquer'' approach using continually refined grids can face
potential difficulty.  For $60^3$ grid points, the near-grid method
underestimates the Bader charge by 0.01e, while the weight method
underestimates it by 0.005e, again showing faster convergence.

\section{Computational effort}
\label{sec:comput}
The weight method is computationally efficient, requiring overall effort
that scales linearly with the number of grid points. The total computer
time is comprised of two primary tasks: the sorting of charge density costs
$O(N \log N)$ with $N$ grid points, and the atomic weight evaluation on the
sorted grid points beginning from grid point with maximum density requires
at most $N \x N_\text{atom}$ computer time. The computational effort is
smaller than that, as only the surface grid points which have fractional
atomic weights require $N_\text{atom}$ calculations, while each interior
grid point require only one calculation. Generally, the number of surface
grid points is a small fraction of the number of total grid points, and
scales as $N^{2/3}$. For example, the ratio of the number of surface grid
points to the number of total grid points is $14\%$ in NaCl crystal with
total grid sizes $60^3$.

\begin{figure}[!ht]
\begin{center}
  \includegraphics[width=3in]{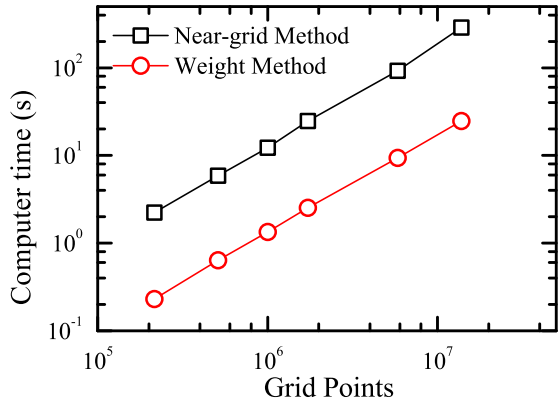}
\end{center}
\caption[Computer time required to analyze the charge density grid for an
eight-atom NaCl cell.]{Computer time required to analyze the charge density
grid for an eight-atom NaCl cell.  The calculations were performed using an
Intel Core2 Quad CPU Q6600, with a clockspeed of 2.40GHz.  The computer
time scales linearly with respect to the number of charge density grid
sizes with the weight method, as with the near-grid method. The weight
method has a smaller prefactor than the near-grid method.}
\label{fig:NaCl_time}
\end{figure}

In a calculation with charge density grid sizes approaching $10^7$--$10^8$
grid points and up to hundreds of atoms in large supercells, we find our
algorithm is not only more accurate, but more efficient than the near-grid
method.  Both methods scale linearly with the number of grid
points. \fig{NaCl_time} shows the linear scaling of computer time required
to analyze the charge density grid for an eight-atom NaCl with the number
of grid points.  The improved efficiency of our algorithm appears to
originate from the lack of a self-consistent refinement of basin
assignment.  Comparing to the near-grid method, which needs refinement
integration, our weight method has small prefactor, although both are
linearly scaled.

\section{Conclusions}
We develop a weight method to integrate functions defined on a discrete
grid over basins of attraction (such as Bader volumes) in an efficient and
accurate manner.  The weight method works with the density on a discrete
grid and assigns volume fractions of the Voronoi cell of each grid point to
surrounding basins.  Starting from the local density maxima, all the grid
points are sorted in density descending order. Grid points can then be
fractionally weighted from the weights of its neighbors with larger
density.  This method depends upon the formulation of flow across that
dividing surfaces between the cells of two neighboring grid points, and can
be applied to uniform or non-uniform grids.

We perform tests on model three-dimensional charge density constructed from
Gaussian functions in FCC cell. The weight method shows that the atomic
integration error is inversely proportional to the 2/3 power of grid
points, while the integration error is inversely proportional to the 1/3
power of grid points using the near-grid method. We also perform tests on
more realistic systems, such as TiO$_2$ bulk and NaCl crystal. In both
cases, the weight method reports maximal atomic integration error at least
one order of magnitude smaller than the near-grid method
systematically. Furthermore, we calculate the Bader charge of NaCl crystal
using these two methods, both give monotonic, and smooth convergence with
respect to the increasing grid sizes, while they converge to slight
different values, by $0.003$ e. The weight method is more accurate than the
near-grid method that require very fine grids.

\begin{acknowledgments}
This research was supported by NSF under grant number DMR-1006077 and
through the Materials Computation Center at UIUC, NSF DMR-0325939, and with
computational resources from NSF/TeraGrid provided by NCSA and TACC.  The
authors thank G.~Henkelman for providing the near-grid code, and for
helpful discussions; and R.~M.~Martin and R.~E.~L.~Deville for helpful
discussions.
\end{acknowledgments}

\appendix
\section{Quadratic error in one-dimension}
\label{sec:quad}
The weight method for integration of the Bader charge volume has error that
is quadratic in the grid spacing in one dimension.  Consider the charge
density $\rho(x)$ evaluated on a regular grid with spacing $\grid$.  In
\eqn{integrate}, there are only two grid points where $\wXA$ is not
exactly 0 or 1; these are the boundary points, and each is adjacent to a
fixed point.  In one dimension, the contributions to \eqn{integrate} that
could produce errors linear in $\grid$ only come from those points; the
integration of the interior produces a total error that is quadratic in
$\grid$.  Hence, without loss of generality, we consider a single boundary
point, and show that its contribution to \eqn{integrate} produces an error
that is of the order $\grid^2$, rather than $\grid$.

Let $X$ be an boundary point, where the basin $A$ lies to its left.  This
requires that $\rho(X-\grid) > \rho(X)$, and
$\rho(X+2\grid)>\rho(X+\grid)$.  Finally, in order for $\wXA$ to not be
identically 1, $\rho(X)<\rho(X+\grid)$.  This means that there is a point
$X+\delta$ for $\delta\in[0,\grid]$ such that $\rho'(X+\delta)=0$.
Then, the flux from \eqn{flux} is
\beq
J_{X\to X-\grid} = \frac{\rho(X-\grid) - \rho(X)}%
{\rho(X-\grid) + \rho(X+\grid) - 2\rho(X)}
\label{eqn:1Dflux}
\eeq
and $J_{X\to X+\grid} = 1-J_{X\to X-\grid}$; finally, as
$\rho(X)<\rho(X+\grid)$, $J_{X+\grid\to X}=0$.  Then, $w^A(X-\grid)=1$,
$w^A(X+\grid) = 0$, and so $w^A(X) = J_{X\to X-\grid}$.  Finally, the
contribution to \eqn{integrate} from $X$ is
\beq
\int_{X-\grid/2}^{X+\delta} f(x) dx \approx \grid J_{X\to X-\grid} f(X)
\eeq

To evaluate the integration error, we use a Taylor expansion for $\rho$ and
$f$ around the grid point $X$.  We write $\rho^{(n)} = d^n\rho/dx^n (X)$
and $f^{(n)} = d^nf/dx^n(X)$.  Note that $\rho^{(1)}$ has to scale as
$\grid$ in order for the dividing point $X+\delta$ to lie between $X$ and
$X+\grid$.  The Taylor expansion of $\wXA=J_{X\to X-\grid}$ from
\eqn{1Dflux} to linear order in $\grid$ is
\beq
\wXA \approx \left[\frac12 - \frac{\rho^{(1)}\grid^{-1}}{\rho^{(2)}}
\right] - \grid\frac{\rho^{(3)}}{6\rho^{(2)}} + O(\grid^2)
\eeq
so our integration contribution is
\beq
\grid\left[\frac12 - \frac{\rho^{(1)}\grid^{-1}}{\rho^{(2)}}
\right]f(X) - \grid^2\frac{\rho^{(3)}}{6\rho^{(2)}} f(X) + O(\grid^3).
\label{eqn:1Dweight}
\eeq
To find the true value of the expression, we need to determine $\delta$ to
at least quadratic order in $\grid$; write $\delta = \delta^{(1)}\grid +
\delta^{(2)}\grid^2$, and we have
\beq
\begin{split}
\rho'(X+\delta) =& \rho^{(1)} + \delta\cdot\rho^{(2)} + \frac12 \delta^2\cdot
\rho^{(3)} + O(\grid^3)\\
0 =& \grid (\rho^{(1)}\grid^{-1}) + \grid \delta^{(1)}\rho^{(2)}\\
&+\grid^2\delta^{(2)}\rho^{(2)} + \frac12 \grid^2
\left(\delta^{(1)}\right)^2 \rho^{(3)} + O(\grid^3)
\end{split}
\eeq
which is solved by
\beq
\begin{split}
\delta^{(1)} &= -\frac{\rho^{(1)}\grid^{-1}}{\rho^{(2)}}\\
\delta^{(2)} &= -\frac{\left(\rho^{(1)}\grid^{-1}\right)^2 \rho^{(3)}}%
{2\left(\rho^{(2)}\right)^3}.
\end{split}
\eeq
With our quadratic approximation for $\delta$, we can integrate $f(x)$ as
\beq
\begin{split}
\int_{X-\grid/2}^{X+\delta} f(x) dx 
=& \left.(x-X)f(X) + \frac12 (x-X)^2f^{(1)} +
O((x-X)^3)\right|_{X-\grid/2}^{X+\delta} \\
=& \grid\left[\frac12 - \frac{\rho^{(1)}\grid^{-1}}{\rho^{(2)}}
\right]f(X)\\
&+ 
\frac{\grid^2}{2}\left(\frac{\rho^{(1)}\grid^{-1}}{\rho^{(2)}}\right)^2
\left[f^{(1)} - f(X)\frac{\rho^{(3)}}{\rho^{(2)}}\right] + O(\grid^3),
\end{split}
\eeq
which agrees with the contribution from our weight integration in
\eqn{1Dweight} up to an error of order $\grid^2$.  As a special case,
consider $f(x)=1/\grid$; then the integral 
\beq
\frac1\grid \int_{X-\grid/2}^{X+\delta} dx = \wXA + O(\grid)
\eeq
which shows that the weight is the volume fraction of the Voronoi volume
belonging to basin $A$ to first order in $\grid$.

\end{document}